# *α*-In$_2$Se$_3$ based Ferroelectric-Semiconductor Metal Junction for Non-Volatile Memories

Atanu K. Saha[1], Mengwei Si[1,2], Peide Ye[1,2], Sumeet K. Gupta[1]

[1]*School of Electrical and Computer Engineering, Purdue University, West Lafayette, IN 47907, US*

[2]*Birck Nanotechnology Center, West Lafayette, IN 47907, US*

In this work, we theoretically and experimentally investigate the working principle and non-volatile memory (NVM) functionality of 2D α-In$_2$Se$_3$ based ferroelectric-semiconductor-metal-junction (FeSMJ). First, we analyze the semiconducting and ferroelectric properties of α-In$_2$Se$_3$ van-der-Waals (vdW) stack via experimental characterization and first-principle simulations. Then, we develop a FeSMJ device simulation framework by self-consistently solving Landau-Ginzburg-Devonshire (LGD) equation, Poisson's equation, and charge-transport equations. Based on the extracted FeS parameters, our simulation results show good agreement with the experimental characteristics of our fabricated α-In$_2$Se$_3$ based FeSMJ. Our analysis suggests that the vdW gap between the metal and FeS plays a key role to provide FeS polarization-dependent modulation of Schottky barrier heights. Further, we show that the thickness scaling of FeS leads to a reduction in read/write voltage and an increase in distinguishability. Array-level analysis of FeSMJ NVM suggests a 5.47x increase in sense margin, 18.18x reduction in area and lower read-write power with respect to Fe insulator tunnel junction (FTJ).

Ferroelectric (Fe) materials have gained an immense research interest for their applications in electronic[1-3] and optoelectronic[4] devices due to their electrically switchable spontaneous polarization and hysteretic characteristics. Fe materials with high bandgap, called Fe-insulators, have been extensively investigated for versatile non-volatile memory (NVM) devices, such as Fe-random-access memory (Fe-RAM)[5], Fe-field-effect-transistor (Fe-FET)[6-7], Fe-tunnel-Junction (FTJ)[8-10], etc. Unlike Fe-RAM and Fe-FET, where the Fe layer acts as a capacitive element, the FTJ functionality depends on the tunneling current through the Fe layer. In the FTJ, the Fe layer is sandwiched by two different metal electrodes. Due to the different properties (e.g. the screening length) of the metal electrodes, the tunneling barrier height at the metal-Fe interface of FTJ depends on the polarization direction. Thus, FTJ can exhibit polarization-dependent tunneling-resistance that facilitates the sensing of its polarization, leading to the design of a two-terminal NVM element[8]. However, as the dominant transport mechanism of FTJ is direct tunneling, to obtain a desired current density for sufficient operational speed, the Fe-insulator thickness needs to be significantly low (<3nm for HZO[9]). Unfortunately, with thickness scaling, the polarization of the Fe-insulator decreases[9], which reduces the ratio of the tunneling-resistance[9]. Therefore, the distinguishability of the FTJ memory states decreases with the thickness scaling. In addition, most of the Fe-insulators (BTO, BFO, doped-HfO$_2$) comprise of oxygen atoms and the dynamic change in oxygen vacancies can play a major role in its ferroelectric characteristics[11]. Therefore, a decrease in ferroelectricity with scaling and issues related to oxygen vacancies lead to significant challenges in the design and implementation of FTJ-based NVMs.

Similar to Fe-insulator, Fe material with low bandgap called Fe-semiconductor (FeS) also exhibit spontaneous polarization which is switchable via applied electric-field[12-16]. Among different FeS materials, the van-der-Waals (vdW) stack of α-In$_2$Se$_3$ has recently been discovered as a 2D FeS material that can retain the ferroelectric and semiconducting properties even for a monolayer thickness[12-16]. This suggests a remarkable possibility for thickness scaling. In addition, as α-In$_2$Se$_3$ is not an oxide, the issues related to oxygen vacancies in oxide Fe-insulator are expected to be non-existent in this FeS material. Recently, similar to FTJ, a metal-FeS-metal junction device (called FeSMJ) has been demonstrated to exhibit polarization-dependent resistance states[17]. Unlike FTJ, FeSMJ can provide significant current density even with a high FeS thickness and it does not require asymmetric metal electrodes for NVM functionalities[17]. To understand such unique working principle of FeSMJ devices and to enable their device-level optimization, a detailed analysis of the material properties α-In$_2$Se$_3$ as well as the device characteristics of FeSMJ, is needed. To address this need, in this work, we experimentally and theoretically analyze α-In$_2$Se$_3$ based FeSMJ devices and examine their thickness scalability. Our analysis is based on experimental characterization and first-principle simulations of α-In$_2$Se$_3$ vdW stack, and experimental measurement and self-consistent device simulation of the FeSMJ devices. Finally, we investigate the thickness scalability of FeSMJ and compare it with FTJ at the device and array levels to analyze its potential for NVM applications.

To begin with, we first discuss the material properties of α-In$_2$Se$_3$. Unit cells of the α-In$_2$Se$_3$ monolayer are shown in Fig.1(a-d) indicating non-centrosymmetric crystal structure, where the central Selenium (Se) atom is spontaneously displaced from the centrosymmetric position. As a result, α-In$_2$Se$_3$ exhibits both in-plane (Fig. 1(a-b)) and out-of-plane (Fig. 1(c-d)) spontaneous polarization ($P_{XY}$ and $P_Z$, respectively). The arrangement of α-In$_2$Se$_3$ layers in a vdW stack is shown in Fig. 1(e) where each layer is separated by ~0.3nm of vdW gap (vacuum region)[12-14]. Employing the vdW stack of α-In$_2$Se$_3$ as the FeS layer in a metal-FeS-metal configuration, the FeSMJ structure is shown in Fig. 1(f). Now, to characterize its semiconducting and ferroelectric properties, α-In$_2$Se$_3$ vdW stack was grown by the melt method with a layered non-centrosymmetric rhombohedral R3m structure[16]. The details of the fabrication process can be found in our previous work ref. **16**. The high-angle annular dark-field STEM (HAADF-STEM) image of thin α-In$_2$Se$_3$ flake is shown in Fig. 1(g) that signifies a high-quality single-crystalline hexagonal structure. The photoluminescence measurement (Fig. 1(h)) of bulk α-In$_2$Se$_3$ suggests a direct optical/direct bandgap of ~1.39eV. To analyze the semiconducting properties further, we conduct first-principle simulations (based on density function theory (DFT)) in Quantum Espresso (QE)[18-19] with hybrid orbital (HSE) functional correction[20]. Unlike previous simulation-based studies of mono-layer or a stack with a few numbers of layers[12], we consider bulk α-In$_2$Se$_3$ vdW stack by taking a super-cell of three In$_2$Se$_3$ layers (Fig. 1(i)) periodically repeated in all three directions (along *x*, *y* and *z*-axis). Note that the thickness of our experimental sample is 120nm (that approximately exhibits 120 layers), therefore, we investigate the bulk properties, rather than a few-layer system. The simulated energy-dispersion (*E-k*) relation is shown in Fig. 1(j) that illustrates a direct optical gap of ~1.4eV (consistent with experiments) and an indirect bandgap of ~1.3eV. Similarly, the density of states, *D(E)* is shown in Fig. 1(k) that suggests a lower density of conduction state compared to the valence states. Hence, the equilibrium Fermi level ($E_F$) (for an undoped α-In$_2$Se$_3$) is closer to the conduction band minima ($E_C$) compared to the valence band maxima ($E_V$). We utilize this $D(E)$ characteristics in our FeSMJ device simulation for the calculation of mobile carrier concentration in the FeS layers, as discussed subsequently.

Next, we analyze the ferroelectric properties of the α-In$_2$Se$_3$ vdW stack. Fig. 2(a) shows the piezoresponse force microscopy (PFM) phase versus applied voltage hysteresis loop of a 120 nm thick α-In$_2$Se$_3$ stack that suggests a ferroelectric polarization switching with a coercive voltage of ~2V. However, due to the semiconducting properties of α-In$_2$Se$_3$, a direct measurement of spontaneous polarization through conventional measurement of polarization-voltage characteristics is not possible[16]. Hence, we perform the Berry phase analysis (based on the modern theory of polarization)[21] on the DFT wave-functions of α-In$_2$Se$_3$ in QE. Our analysis suggests an out-of-plane remanent polarization ($P_R$) of ~7.68$\mu$C/cm$^2$. Note that, due to the periodic nature of the bulk vdW stacks, the calculated $P_R$ for bulk α-In$_2$Se$_3$ (by the Berry phase method) is higher in magnitude compared to the previously calculated $P_R$ (by dipole correction method) for a few-layer system[12]. To further understand the ferroelectric properties of α-In$_2$Se$_3$, the microscopic potential energy (averaged across the x-y plane) along the FeS thickness (z-axis) obtained from the DFT simulation is shown in Fig. 2(b). The extracted macroscopic potential (Fig. 2(c)) suggests an opposite electric-field in FeS layers and vdW gaps. Now, the electrostatic condition that needs to be satisfied at the interface of FeS and vdW gap can be written as,

$$\epsilon_0 E_{vdW} = \epsilon_0 \epsilon_r E_{FeS} + P \qquad (1)$$

Here, $E_{vdW}$ and $E_{FeS}$ are the E-fields in the vdW gap and FeS layer, respectively; $\epsilon_r$ is the relative background permittivity of the FeS layer, $\epsilon_0$ is the vacuum permittivity and $P$ is the spontaneous polarization. The above equation suggests that, $E_{vdW}$ and $E_{FeS}$ can be non-zero and hold the opposite sign if and only if the $P$ is non-zero. Therefore, the opposite electric-field in FeS layers and vdW gaps confirms the existence of spontaneous polarization in the FeS layer. Using the calculated value of $P$, $E_{vdW}$ and $E_{FeS}$, we obtain $\epsilon_r$=~7 for the FeS layer from Eqn. (1). Further, we calculate the total energy with respect to the change in $P$ based on the nudge-elastic-band (NEB)[22] method in QE. The change in $P$ is captured by sweeping the position of the central Se atom gradually from one stable position to another stable position as shown in Fig. 2(d) followed by performing Berry phase calculation for $P$ in each of the intermediate states. To capture the finite temperature effect in total energy ($u$), we have considered phonon-energy correction[23] for 300K temperature. The resultant $u$-$P$ characteristics are shown in Fig. 2(d) signifying a double-well energy landscape with a barrier between the two stable polarization states. We fit the simulated $u$-$P$ characteristics with Landau's free energy polynomial[24] as shown in Fig. 2(d) based on the following equation.

$$u = \tfrac{1}{2}\alpha P^2 + \tfrac{1}{4}\beta P^4 + \tfrac{1}{6}\gamma P^6 + \tfrac{1}{8}\delta P^8 \qquad (2)$$

The obtained Landau coefficients ($\alpha$, $\beta$, $\gamma$ and $\delta$) are shown in the inset of Fig. 3(d). Here, a non-zero coefficient for the 8$^{th}$ order term ($\delta$) is required to capture the flattened shape of the energy barrier. Based on the extracted parameters that correspond to the ferroelectric and semiconducting properties of α-In$_2$Se$_3$, we self-consistently solve Landau-Ginzburg-Devonshire (LGD) equation[3], Poisson's equation and semiconductor charge equations for the FeSMJ structure. Then, we use the potential profile in a NEGF framework to calculate the current in the FeSMJ assuming ballistic transport in the thickness direction. The simulation flow is presented in Fig. 3(a) showing the utilization of the extracted parameters for the device simulation. In our simulation, we consider the vdW gap of 3.3Å between the subsequent FeS layers (obtained from DFT simulation with structural relaxation) along with a vdW gap of 1.65Å between the metal and FeS layer as shown in Fig. 3(b). We utilize this simulation framework along with the experimental results to investigate the FeSMJ device characteristics.

The top-view of the fabricated FeSMJ device is shown as a false-color SEM image in Fig. 4(a). Here, the FeS thickness ($T_{FeS}$) is 120nm and the same metal (Nickel (Ni)) is used as the top and bottom metal contacts. The measured current ($I$) vs voltage ($V$) characteristics (Fig. 4(b)) exhibit a counter-clockwise hysteresis due to which the FeSMJ shows two different conductive or resistive states. Let us define the current at low-resistance state (LRS) and high-resistance state (HRS) as $I_{LRS}$ and $I_{HRS}$, respectively. Here, one noticeable thing is that the characteristics are asymmetric with respect to the voltage polarity. For example, the hysteresis window, currents ($I_{LRS}$ and $I_{HRS}$), and their ratio ($I_{LRS}/I_{HRS}$) are unequal for positive and negative applied voltages. To understand the possible origin of asymmetry as well as the device operation, we perform device-level simulation. The simulated $I$-$V$ curve considering $T_{FeS}$=120nm and Ni as metal contacts is shown in Fig. 4(b) indicating a good agreement with the experimental results. Due to a Schottky barrier at the metal-FeS interface (Fig. 3(b)), the observed current in the FeSMJ is due to the electron injection from the metal to FeS via Schottky tunneling along with direct tunneling through the vdW gaps.

Now, to understand the working principle of FeSMJ, the equilibrium band diagram of the device (along the $z$-axis i.e FeS thickness) is shown in Fig. 4(c)-i. Note, the band diagram is for an undoped α-In$_2$Se$_3$, in which the Fermi level ($E_F$) is closer to the conduction band, as discussed before. Without any loss of generality, let us assume that initially, all the FeS layers are -$z$ directed polarization (-$P$). Let us call the left electrode M1 and right electrode M2. Now, the polarization-induced negative (positive) bound charges appear in the FeS near the M1 (M2) interface. The bound charges and the work-function difference between the metal and FeS induce an E-field within the vdW gap and the FeS layers. As a result, holes (electrons) appear at FeS-M1(M2) interface to partially compensate the negative (positive) bound charges. At the same time, a built-in potential with opposite polarity appears across the two FeS-M junctions, yielding different Schottky barrier height ($\phi_B$) for the mobile carriers (electrons/holes). For example, in Fig. 4(c)-i, $\phi_B$ at the FeS-M1 interface is higher than the FeS-M2 interface due to the negative and positive voltage across the respective vdW gaps, which corresponds to positive and negative bound charges, respectively. Depending on whether the electron-injecting barrier exhibits low or high $\phi_B$, FeSMJ operates in LRS or HRS. Moreover, voltage-driven polarization switching can enable transitioning between LRS and HRS, and vice versa. To understand this, let us consider a positive bias at M2. Hence, the electron injection takes place from M1 to FeS. As the corresponding $\phi_B$ is high, FeSMJ operates in HRS. At the same time, hole (electron) concentration at the FeS-M1 (M2) interface increases. As the hole DOS is higher compared to the electron DOS, the increase in hole concentration is higher compared to the electron concentration (Fig. 4(c)-ii). This leads to a higher electric-field near the FeS-M1 interface compared to the FeS-M2 interface. At sufficiently high positive voltages (~2V in Fig. 5d-ii), the E-field near the FeS-M1 increases beyond the coercive field. Hence, a few layers near the M1 interface switch to +$P$ (+$z$ directed) as shown in Fig. 4(c)-iii. Consequently, $\phi_B$ at the FeS-M1 interface significantly decreases, leading to an abrupt increase in the current (LRS). The LRS operation continues even when the voltage is reduced to 0 due to $P$ retention. Now, when a negative voltage is applied at M2, electron injects from M2 to FeS (Fig. 4(c)-iv). As the corresponding $\phi_B$ is low, the FeSMJ continues to operate in LRS. With further increase in the negative polarity voltage, the electric-field near the M1-FeS interface (which is higher in magnitude than M2-FeS, as explained before) switches the polarization back to –$P$ (Fig. 4(c)-vi). This significantly reduces the E-field near the M1-FeS interface, the effect of which penetrates throughout the FeS including near the electron injecting electrode (FeS-M2 interface). This reduces the current. However, unlike $V$>0V, where switching from HRS to LRS is abrupt, here the change in current is gradual. This is because for $V$>0V, the change in $\phi_B$ of the electron injecting junction (FeS-M1) is large due to $P$-switching near that interface. On the other hand, for $V$<0V, $\phi_B$ of the electron injecting electrode (FeS-M2) does not change significantly as $P$-switching occurs on the other electrode. Therefore, the voltage hysteresis, currents ($I_{LRS}$ and $I_{HRS}$) and their ratio ($I_{LRS}/I_{HRS}$) are asymmetric with respect to the voltage polarity (i.e. lower for $V$<0 than for $V$>0). To complete the discussion, if the initial $P$ is opposite (+$P$ for all the FeS layers), the $I$-$V$ characteristics would be the opposite of what we discussed so far, i.e. a gradual HRS-to-LRS switching for $V$>0V (for +$P$ to -$P$ switching near the FeS-M2 interface causing a non-significant change in the electron injecting $\phi_B$) and an abrupt LRS-to-HRS switching for $V$<0V (due to -$P$ to +$P$ switching near the FeS-M2 interface causing a change in the electron injecting $\phi_B$). In summary, the appearance of mobile carriers leads to a non-uniform electric field in the FeS layers that leads to partial (and non-homogeneous) $P$-switching in the FeS yielding asymmetric $I$-$V$ characteristics in FeSMJ.

So far, we have discussed the FeSMJ characteristics for $T_{FeS}$ = 120nm. Interestingly, in scaled FeSMJ (i.e. $T_{FeS}$ = 60nm), the coercive field can be achieved at a lower voltage and before the appearance of a significant density of mobile carriers. Due to low mobile carrier concentrations, the electric-field in FeS layers becomes less non-uniform (compared to $T_{FeS}$ = 120nm) and that leads to a complete polarization switching along with the FeS thickness. Therefore, the resultant $I$-$V$ characteristics for $T_{FeS}$ = 60nm show significantly less asymmetric behavior (Fig. 5(a)) in terms of hysteresis, current magnitude ($I_{LRS}$ and $I_{HRS}$) and their ratio ($I_{LRS}/I_{HRS}$). Such a unique FeS thickness scaling behavior now brings us to the analysis of the effect of FeS thickness scaling on the FeSMJ $I$-$V$ characteristics, which offers useful insights into the device optimization in the context of NVM applications.

The simulated $I$-$V$ characteristics of FeSMJ with different $T_{FeS}$ are shown in Fig. 5(a). With thickness scaling, the coercive field can be achieved at a lower applied voltage, therefore, the required voltage to switch the resistance state (called write voltage, $V_{write}$) decreases as shown in Fig. 5(b). At the same time, a decrease in $T_{FeS}$ leads to an increase in electric-field (for the same applied voltage) yielding an increase in both $I_{LRS}$ and $I_{HRS}$ as shown in Fig. 5(a). Recall that the mobile carrier concentration in FeS tends to (partially) compensate for the effect of polarization-induced bound charge. As the mobile-carrier concentration decreases with the decrease in $T_{FeS}$ (as discussed above), therefore, the effect of polarization-induced bound charge becomes more prominent. Hence, the polarization-dependent modulation in $\Phi_B$ increases with the decreases in $T_{FeS}$, which leads to an increase in $I_{LRS}/I_{HRS}$ (Fig. 5(b)). Consequently, improved distinguishability ($I_{LRS}/I_{HRS}$) along with low voltage NVM operation can be achieved by scaling down the FeS thickness.

Next, we evaluate the FeSMJ NVM performance in comparison with FTJ. In an array, each NVM cell is comprised of a FeSMJ/FTJ connected in series with an access transistor (Fig. 5(c)). In this analysis, we assume a 10nm finfet (with 4 fins) as the access transistor. For FeSMJ, we consider $T_{FeS}$=60nm and for FTJ, we consider a $HfO_2$/$SiO_2$ (4nm/0.4nm) based device with Al and doped-Si as contacts from ref. 9. The device-level comparison (Fig. 5(d)) suggests that the FeSMJ has a key advantage of higher current density than FTJ (due to Schottky transport in the former as opposed to direct tunneling in the latter), while the distinguishability between the states ($I_{LRS}/I_{HRS}$ ~$10^3$) are similar for both. This enables more aggressive area scaling in FeSMJ compared to FTJ. Considering a layout design by following ref. 25 for an array size of 1MB, our analysis suggests a 0.87x lower read power, 0.94x lower write power and 5x higher sense margin with an 18.18x less area for FeSMJ compared to FTJ (Fig. 5(e)). Due to such remarkable benefit of FeSMJ over FTJ for NVM application, further exploration of FeSMJ is required to investigate its retention characteristics in addition to its correlation with scaling.

In summary, we show that the spontaneous FeS polarization induces a built-in potential across the vdW gap between FeS and metal contact leading to a polarization-dependent Schottky barrier for electron injection. By switching the polarization via applied voltage, the Schottky barrier height can be modulated, which leads to transitions between HRS and LRS in FeSMJ. Further, we show that in FeSMJ with high FeS thickness, the appearance of mobile carriers can lead to a partial polarization switching in FeS yielding asymmetric *I-V* characteristics. However, with thickness scaling, the asymmetry reduces due to complete polarization switching and at the same time $I_{LRS}/I_{HRS}$ increases. Most importantly, FeSMJ exhibits a significantly higher current density due to Schottky tunneling compared to FTJ. Due to such appealing characteristics of FeSMJ and fundamental differences in the transport mechanisms compared to FTJ, FeSMJ NVM array exhibits significantly higher energy efficiency and integration density than FTJ NVM.

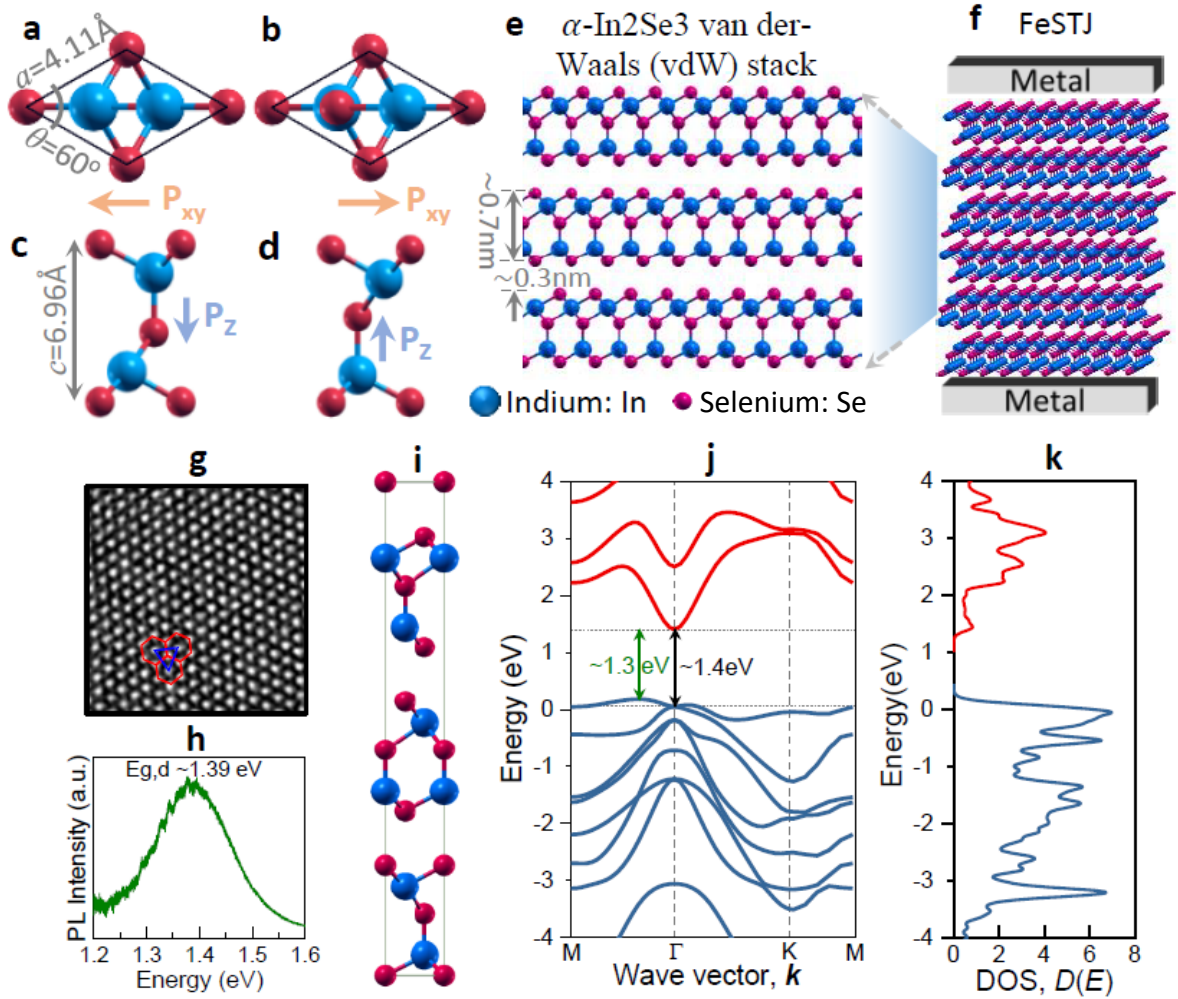

Fig. 1: Unit cell of α-In$_2$Se$_3$ crystal (a-b) top view and (c-d) side view showing different in-plane and out-of-plane polarization directions, respectively. (e) van-der-Waals (vdW) stack of α-In$_2$Se$_3$ ferroelectric semiconductor (FeS) and (f) FeSMJ device structure. (g) STEM image of the fabricated bulk α-In$_2$Se$_3$ surface. (h) Measured photoluminescence (PL) spectrum showing a direct optical/direct bandgap of ~1.39eV. (i) supercell of bulk α-In$_2$Se$_3$ vdW stack. (j) Energy-dispersion (E-k) relation and (k) density of states (D(E)) of α-In$_2$Se$_3$ vdW stack from DFT simulation.

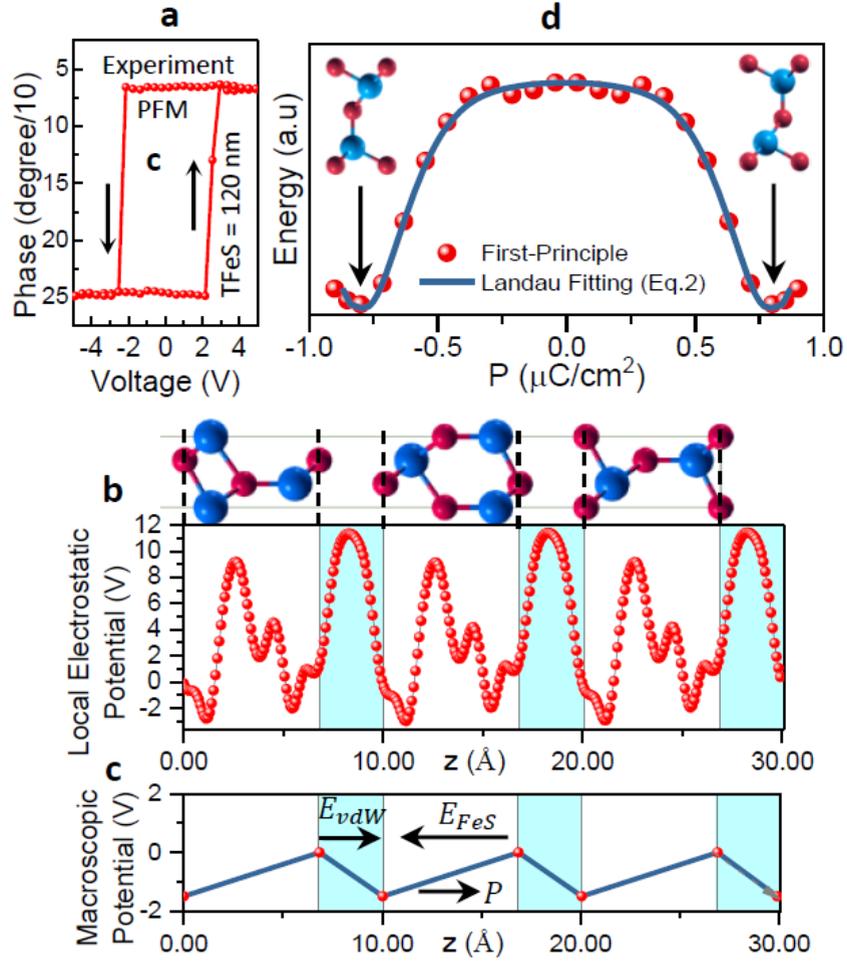

Fig. 2: (a) Measured PFM phase response of FeS with $T_{FeS}$=120nm. (b) Local electrostatic potential and (c) macroscopic potential profile in $\alpha$-In$_2$Se$_3$ vdW stack (along the z-axis) showing the opposite E-field profile in the vdW gap (vacuum) and In$_2$Se$_3$ layers. (d) Polarization ($P$) vs energy ($u$) landscape of $\alpha$-In$_2$Se$_3$ from DFT+NEB (nudget-elastic-band) simulation.

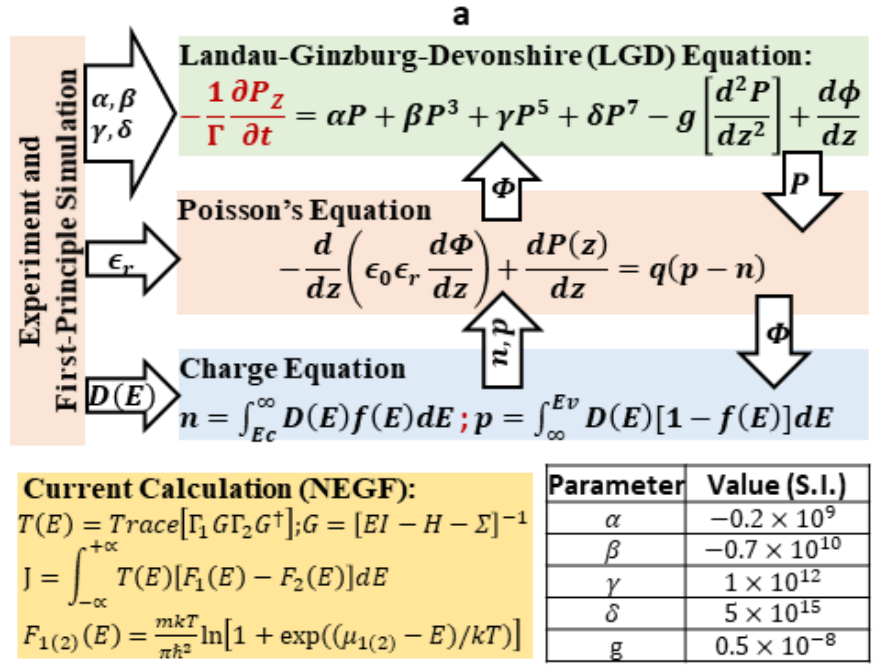

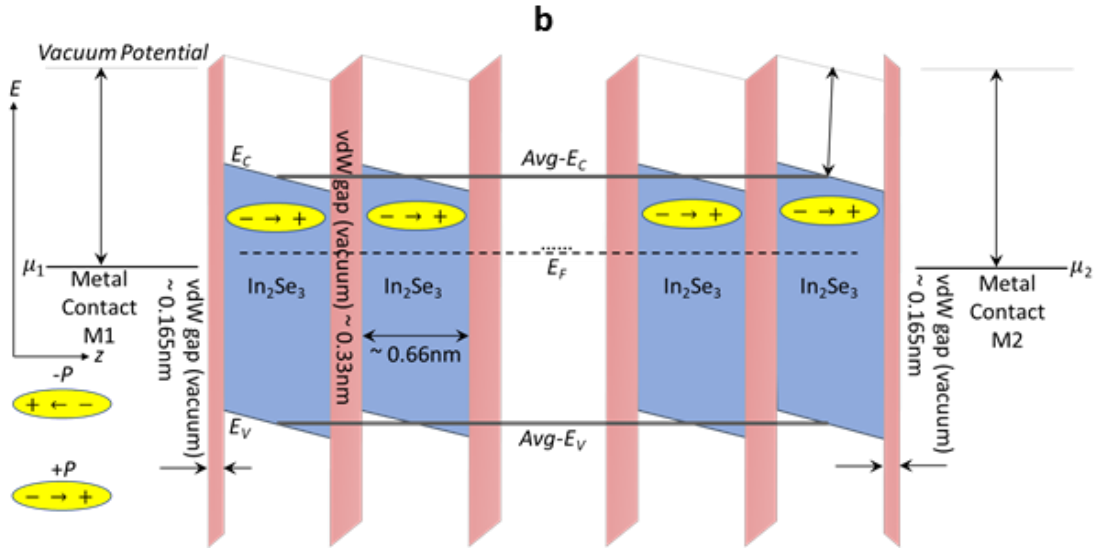

Fig. 3: (a) Self-consistent simulation flow (Poisson + Charge + LGD) and equations used for FeSMJ device simulation and table showing parameters used in the simulation. (b) Band alignment of M-FeS-M structure before equilibrium.

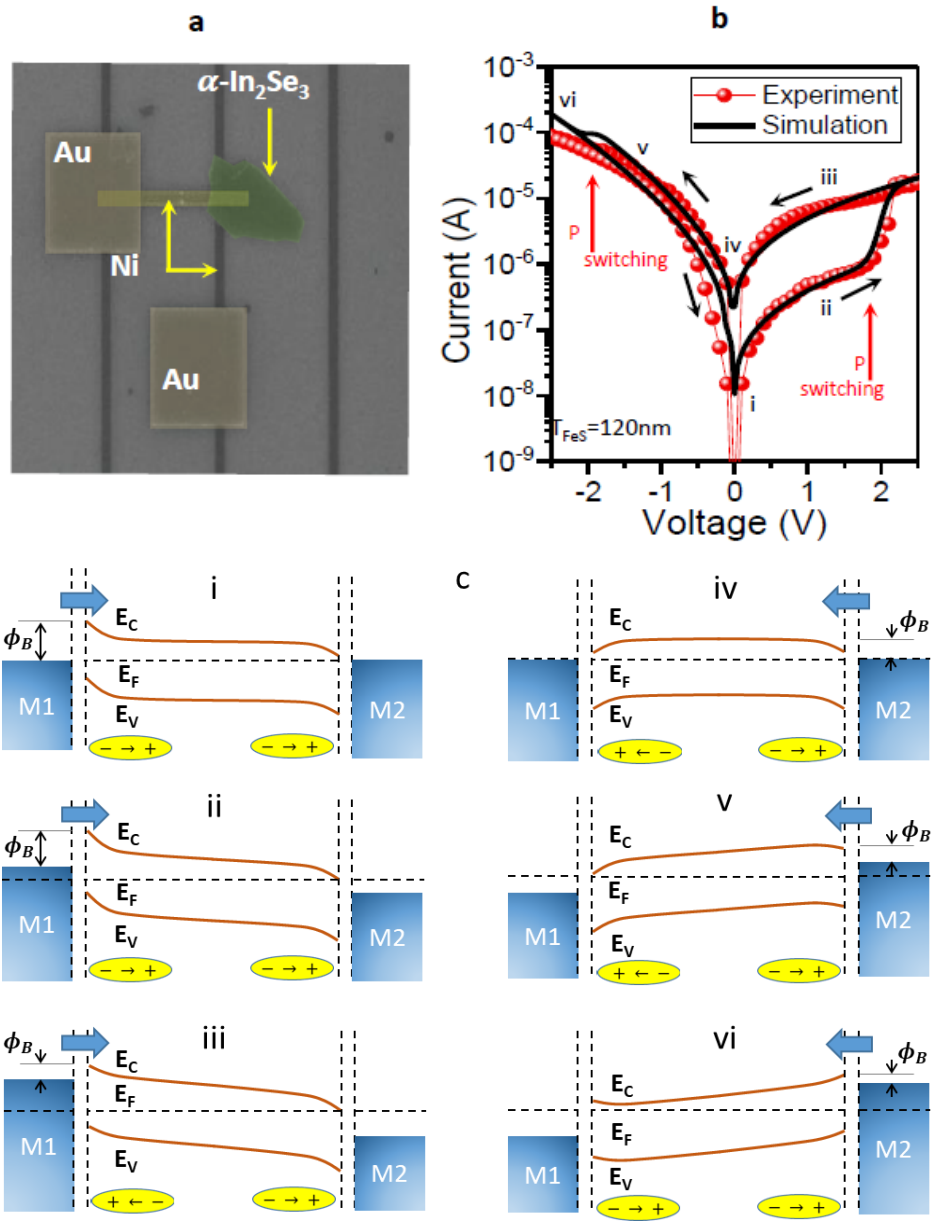

Fig. 4: (a) (a) SEM image of the fabricated FeSMJ. (b) Measured and simulated *I-V* characteristics of FeSMJ with FeS thickness, $T_{FeS}$=120nm. (c) Band diagram of FeSMJ that corresponds to the different points marked in (b). Here, $E_C$ and $E_V$ are taken at the center of each FeS layers and the vdW regions within the FeS layers are not shown for the clarity.

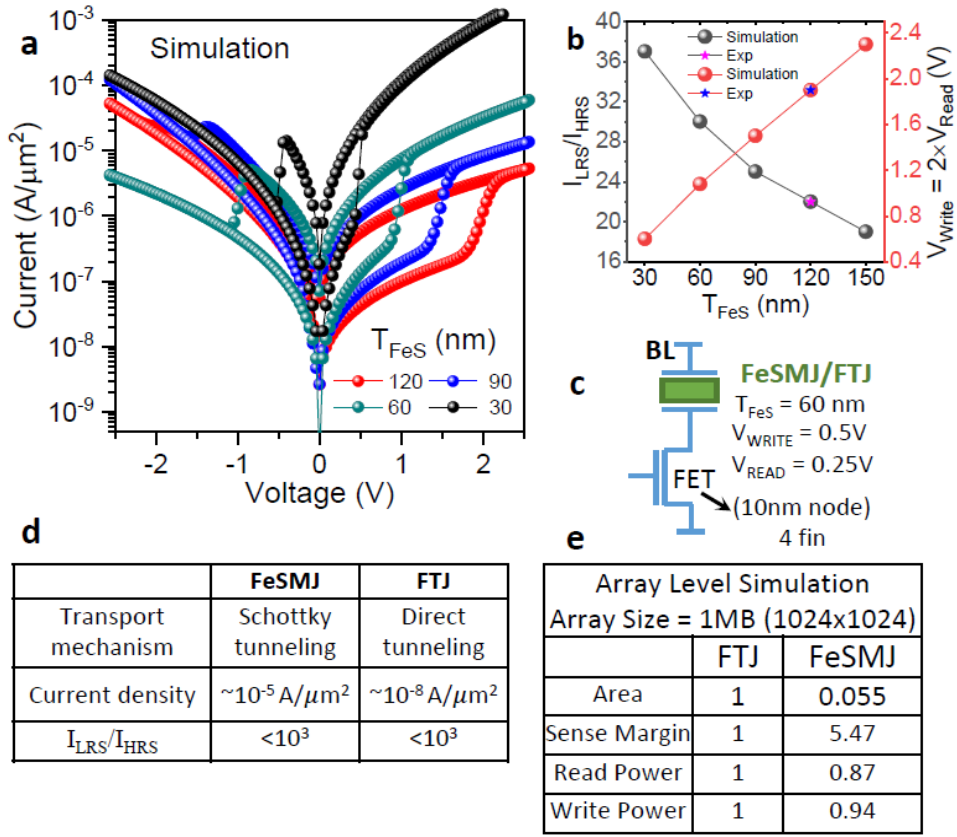

Fig. 5: (a) *I-V* characteristics, (b) write voltage ($V_{Write}$), read voltage ($V_{Read}$) and $I_{LRS}/I_{HRS}$ of FeSMJ for different FeS thickness. (c) FeSMJ/FTJ memory cell (1T-1R). (d) device and (e) array level comparison between FeSMJ and FTJ.